\documentclass[multphys]{svmult}

\usepackage{makeidx}         
\usepackage{graphicx}        
\usepackage{multicol}        
\usepackage[bottom]{footmisc}

\makeindex             

\begin{document}

\title*{Young massive star clusters in the era of the {\sl Hubble
Space Telescope}}


\author{Richard de Grijs\inst{1,2}}


\institute{Department of Physics \& Astronomy, University of
Sheffield, Hicks Building, Hounsfield Road, Sheffield S3 7RH, UK
\texttt{R.deGrijs@sheffield.ac.uk}
\and 
National Astronomical Observatories, Chinese Academy of Sciences, 20A
Datun Road, Chaoyang District, Beijing 100012, China}

%
%
\maketitle

\begin{abstract}
The {\sl Hubble Space Telescope (HST)} has been instrumental in the
discovery of large numbers of extragalactic young massive star
clusters (YMCs), often assumed to be proto-globular clusters (GCs). As
a consequence, the field of YMC formation and evolution is thriving,
generating major breakthroughs as well as controversies on annual (or
shorter) time-scales. Here, I review the long-term survival chances of
YMCs, hallmarks of intense starburst episodes often associated with
violent galaxy interactions. In the absence of significant external
perturbations, the key factor determining a cluster's long-term
survival chances is the shape of its stellar initial mass function
(IMF). It is, however, not straightforward to assess the IMF shape in
unresolved extragalactic YMCs. I also discuss the latest progress in
worldwide efforts to better understand the evolution of entire cluster
populations, predominantly based on {\sl HST} observations, and
conclude that there is an increasing body of evidence that GC
formation appears to be continuing until today; their long-term
evolution crucially depends on their environmental conditions,
however.
\end{abstract}

\section{Introduction}
\label{intro.sec}

Young, massive star clusters (YMCs) are the hallmarks of violent
star-forming episodes triggered by galaxy collisions and close
encounters. This field has seen major progress and a flurry of renewed
interest ever since such YMCs were first reported in the starburst
galaxy NGC 1275 by Holtzman et al. (1992) using pre-COSTAR {\sl Hubble
Space Telescope (HST)} images. The question remains, however, whether
or not at least a fraction of the compact YMCs seen in abundance in
extragalactic starbursts, are potentially the progenitors of ($\ge 10$
Gyr) old globular cluster (GC)-type objects -- although of higher
metallicity than the present-day GCs. If we could settle this issue
convincingly, one way or the other, such a result would have
far-reaching implications for a wide range of astrophysical questions,
including our understanding of the process of galaxy formation and
assembly, and the process and conditions required for star (cluster)
formation. Because of the lack of a statistically significant sample
of YMCs in the Local Group, however, we need to resort to either
statistical arguments or to the painstaking approach of case-by-case
studies of individual objects in more distant galaxies.

\section{Individual YMC evolution}

The evolution to old age of young clusters depends crucially on their
stellar initial mass function (IMF). If the IMF slope is too shallow,
i.e., if the clusters are significantly deficient in low-mass stars
compared to, e.g., the solar neighbourhood, they will likely disperse
within about a Gyr of their formation (e.g., Chernoff \& Shapiro 1987;
Chernoff \& Weinberg 1990; Goodwin 1997b; Smith \& Gallagher 2001;
Mengel et al. 2002). As a case in point, Goodwin (1997b) simulated the
evolution of $\sim 10^4 - 10^5 M_\odot$ YMCs similar to those observed
in the Large Magellanic Cloud (LMC), with IMF slopes $\alpha = 2.35$
(Salpeter 1955; where the IMF is characterised as $\phi(m_\ast)
\propto m_\ast^{-\alpha}$, as a function of stellar mass, $m_\ast$)
and $\alpha = 1.50$, i.e., roughly covering the range of (present-day)
mass function slopes observed in LMC clusters at the time he performed
his {\it N}-body simulations. Following Chernoff \& Weinberg (1990),
and based on a detailed comparison between the initial conditions for
the LMC YMCs derived in Goodwin (1997b) and the survival chances of
massive star clusters in a Milky Way-type gravitational potential
(Goodwin 1997a), Goodwin (1997b) concluded that -- for Galactocentric
distances $\ge 12$ kpc -- some of his simulated LMC YMCs should be
capable of surviving for a Hubble time if $\alpha \ge 2$ (or even $\ge
3$; Mengel et al. 2002), but not for shallower IMF slopes for any
reasonable initial conditions (cf. Chernoff \& Shapiro 1987; Chernoff
\& Weinberg 1990). More specifically, Chernoff \& Weinberg (1990) and
Takahashi \& Portegies Zwart (2000), based on numerical cluster
simulations employing the Fokker-Planck approximation, suggest that
the most likely survivors to old age are, additionally, characterised
by King model concentrations, $c \ge 1.0-1.5$. Mengel et al. (2002;
their fig. 9) use these considerations to argue that their sample of
YMCs observed in the Antennae interacting system might survive for at
least a few Gyr, but see de Grijs et al. (2005a), and Bastian \&
Goodwin (2006) and Goodwin \& Bastian (2006), for counterarguments
related to environmental effects and to variations in the clusters'
star-formation efficiencies, respectively.

In addition, YMCs are subject to a variety of internal and external
drivers of cluster disruption. These include internal two-body
relaxation effects, the nature of the stellar velocity distribution
function, the effects of stellar mass segregation, disk and bulge
shocking, and tidal truncation (e.g., Chernoff \& Shapiro 1987; Gnedin
\& Ostriker 1997). All of these act in tandem to accelerate cluster
expansion, thus leading to cluster dissolution -- since expansion will
lead to greater vulnerability to tidally-induced mass loss.

With the ever increasing number of large-aperture ground-based
telescopes equipped with state-of-the-art high-resolution
spectrographs and the wealth of observational data provided by the
{\sl HST}, we may now finally be getting close to resolving the issue
of potential YMC longevity conclusively. To do so, one needs to obtain
(i) high-resolution spectroscopy, in order to obtain dynamical mass
estimates, and (ii) high-resolution imaging to measure their sizes
(and luminosities). One could then construct diagnostic diagrams of
YMC mass-to-light ($M/L$) ratio versus age, and compare the YMC loci
in this diagram with simple stellar population models using a variety
of IMF descriptions.  In de Grijs \& Parmentier (2007; their fig. 2)
we present an updated version of the $M/L$ ratio versus age diagram,
including all of the YMCs for which the required observables are
presently available.

Despite some outstanding issues (particularly for the youngest
clusters), it appears that most of the YMCs for which high-resolution
spectroscopy is available are characterised by ``standard'' Salpeter
(1955) or Kroupa (2001) IMFs. As such, a fraction of the YMCs seen
today may potentially evolve to become old GCs, depending on their
environmental conditions.

\section{The evolution of star cluster systems}

Following the violent relaxation induced by the supernova-driven
expulsion of the left-over star-forming gas, star clusters -- at least
those that survive the ``infant mortality'' phase (i.e., roughly the
first 10--30 Myr of their lives) -- settle back into virial
equilibrium by the time they reach an age of about 40--50 Myr (Bastian
\& Goodwin 2006; Goodwin \& Bastian 2006). Subsequently, the initial
conditions characterising these gas-free bound star clusters are
modified as secular evolution proceeds. Internal (two-body relaxation)
and external effects (due to interactions with the tidal field
associated with the underlying galactic gravitational potential) lead
to tidal stripping and the evaporation of a fraction of the low-mass
cluster stars, thus resulting in the gradual dissolution of star
clusters.

One of the most important diagnostics used to infer the formation
history, and to follow the evolution of a star cluster population is
the CMF (i.e., the number of clusters per constant logarithmic cluster
mass interval, ${\rm d}N/{\rm d}\log m_{\rm cl}$). Of particular
importance is the {\it initial} cluster mass function (ICMF), since
this holds clues to the star and cluster formation processes. The
debate regarding the shape of the ICMF, and of the CMF in general, is
presently very much alive, both observationally and theoretically.
This is so because it bears on the very essence of the star-forming
process, as well as on the formation, assembly history and evolution
of the clusters' host galaxies over cosmic time. Yet, the observable
property one has access to is the CLF (i.e., the number of objects per
unit magnitude, ${\rm d}N/{\rm d}M_V$).

In this respect, a remaining contentious issue is whether the observed
CMF of YMCs will eventually evolve into that of the ubiquitous old
GCs. The GCMF is a Gaussian with a mean $\langle \log(m_{\rm
cl}[M_\odot]) \rangle \sim 5.2-5.3$ and a standard deviation of
$\sigma_{\log m_{\rm cl}} \simeq 0.5-0.6$ dex. It seems to be almost
universal, both {\it among} and {\it within} galaxies. On the other
hand, many CMFs of YMCs appear to be featureless power laws with a
spectral index $\alpha \sim -2$ down to a few $\times 10^3 M_\odot$
(e.g., de Grijs \& Anders 2006; Hunter et al. 2003, for the LMC). Some
cluster systems exhibit differently shaped ICMFs, however (M82 B, de
Grijs et al. 2005b, but see Smith et al., in prep.; NGC 1316,
Goudfrooij et al. 2004; NGC 5253, Cresci et al. 2006). Evolving such
an initial power-law CMF into the near-invariant Gaussian GCMF
regardless both of the host galaxy properties and of the details of
the cluster loci turns out to be most challenging and requires
significant fine-tuning of the models, which is not necessarily
compatible with the available observational constraints (see, e.g.,
Fall \& Zhang 2001 vs. Vesperini et al. 2003; see also Vesperini \&
Zepf 2003).

In order to settle the issues of cluster evolution and ICMF shape more
conclusively, major improvements are required in the near future, both
observationally and theoretically. Observations reaching low-mass
clusters, and with sufficiently accurate photometry, in order to
derive reliable cluster ages, are required to follow the temporal
evolution of the CMF. From a modeling point of view, a better
treatment of the initially loosely bound clusters (i.e., the
low-concentration clusters) is required, since these may account for
the missing link between the Gaussian GCMF and the power laws seen for
YMC systems (Vesperini \& Zepf 2003). In addition, the inclusion of a
time-dependent host galaxy gravitational potential will enable us to
better follow the early evolution of both old GCs and YMCs formed in
interacting and merging galaxies.

%
%
%



\printindex
\end{document}